# The Virtual Observatory and its Benefits for Amateur Astronomers

Petr Škoda

*Astronomical Institute, Academy of Sciences of the Czech Republic, CZ-251 65  Ondřejov, Czech Republic*
*E-mail: skoda@sunstel.asu.cas.cz*

The contemporary astronomical instruments have been producing the unprecedented amount of data. The largest part of this "data avalanche" is being produced by deep all-sky surveys yielding terabytes of raw data per night. Such a great data volumes can hardly even been reduced by automatic pipelines running on supercomputer grids but it is impossible to exploit fully their content by a small group of professional astronomers in the interested research teams.  New tools for collaborative work with heterogeneous data sets spread over distant servers are being developed in the framework of the Virtual Observatory (VO).  As many VO resources are freely  available on the Internet, a new opportunity opens for the amateur astronomers to do professional research using these tools in an  Internet browser on a moderately fast connection.  We give short overview of current and future sky surveys producing data on a millions of targets - hence the term Megasurveys, and we introduce the basic principles of Virtual Observatory and its current applications.

## Megasurveys

The astronomy at the beginning of 21st century is mostly characterized by the avalanche of data being produced by the large CCD mosaics at even small telescopes, hundreds of spectra per exposure on multi-object spectrographs or 3D data cubes from  radio telescope arrays. Considerably higher speed of telemetry and the effective on-board compression makes the current  astronomical satellites the large contributors to data archives as well. There is, however, another source of TB size data sets that is not so obvious - the theoretical models of various physical processes (stellar evolution models, galaxy interaction, cosmological evolution...) run on supercomputers and their distributed networks. Let's have a look at the most important sky surveys that could be of particular interest of the amateur astronomers:

### Sloan Digital Sky Survey (SDSS)

The 2.5m telescope at the Apache point, New Mexico, USA, harnessed with the set of thirty  2kx2k CCD chips (organized in six groups with five colour filters each) and multi-object spectrograph with 640 fibres [http://www.sdss.org/background/telescope.html] has collected in first five years the 5-colour photometry of more than 215 millions objects and about 1.1 million of low resolution spectra (R=3000, 3200-9200 A), among them 150 thousands stars, 675 thousands galaxies, 90 thousands quasars and more than 12 thousands objects of unknown class [http://astro.ncsa.uiuc.edu/catalogs/dposs] .

### DPOSS

Digital Palomar Sky Survey with the extension of Southern sky is the one of the primary sources for creating optical star finding charts [http://astro.ncsa.uiuc.edu/catalogs/dposs] .

### Infrared surveys

The most cited infrared surveys are 2MASS [http://astro.ncsa.uiuc.edu/catalogs/2mass/index.jsp] and DENIS [http://cdsweb.u-strasbg.fr/denis.html]  as a result of the 5-year project of monitoring the sky on 1-m class telescopes from Chile in multiple NIR ranges.

### Microlensing surveys

Specialized, originally microlensing surveys OGLE and MACHO as well as MOA and planet hunting project WASP [http://www.cv.nrao.edu/fits/www/yp_survey.html] are due to its nature the fine source of precise photometric light curves  of many millions of objects.

### Space observatories  and radiotelescopes

Although the archives of well-known satellite observatories like Chandra, HST, Spitzer, ISO, IUE, XMM-NEWTON, INTEGRAL and many others have been already opened or will be opened  in near future to public, the nature of its data and the lack of final science-ready products make them rather cumbersome for the amateurs.  The same concerns the raw radio  telescopes data, although there are already science-ready surveys available.

## The History of Virtual Observatory

To handle effectively the data avalanche, the new research infrastructure had to be established just at the very beginning of 21st century. The astronomers were inspired by facilities common in the research of elementary particles (the technology behind the huge particle accelerators). The key issue there is the decentralization of data storage, joining the computer power in a seamless way using the GRID architecture and providing on-site high level service requiring only exchange of results not raw data  among nodes in distinct research centres [http://www.gridbus.org/papers/WeavingGrid.pdf]. The result of the introduction of the so called Open Grid Service Architecture (OGSA) [http://www.globus.org/alliance/publications/papers/ogsa.pdf] in astronomy lies in the heart of the European leading project of  Virtual Observatory – ASTROGRID [www.astrogrid.org] .

The need of the effective exploration of the SDSS data led in 2000 to the establishing the US founded project called NVO (National Virtual Observatory) [www.us-vo.org] .  In the framework of NVO the basic infrastructure was devised for the queries on distributed servers, cross-matching catalogues and exchanging images of sky at required coordinates. The so called SkyNode servers with copies of SDSS archives were installed worldwide to evenly distribute the requests for archive access from various places [http://skyserver.org/mirrors] .

The Virtual Observatory ideas were met quickly at CDS Strasbourg.  The experience with bibliography and catalogue  databases led to the recognition of  importance of  separation of data itself from meta-data (semantic description of variables). Hence the VOTable standard using the UCDs was established.  Soon the  well-known services like SIMBAD, VIZIER and ALADIN were quickly changed in its back-end part to VO-compatible services allowing better interoperability of CDS databases with external sources in a seamless and scalable way [http://cdsweb.u-strasbg.fr/avo.htx] .

All major astronomical data centres in the world have been adopting the basic VO infrastructures in the newly developed applications and services and soon the international collaboration on VO standards  has been established under the head of  International Virtual Observatory Alliance (IVOA [www.ivoa.net] ). The IVOA organizes IVOA Interoperability meetings  twice a year to assure the standardization of data formats, protocols and query language in all current and future VO applications but its activity is based mostly on the voluntary work.

After six years the VO has presented its capabilities and powerful features in a number of projects in various fields of astronomy, like searching of brown dwarf candidates, discovery of obscured quasars, research of post-AGB stellar evolution, estimating fundamental cosmological parameters and many others.  The amount of data planned to flow from future large telescopes and satellites (LSST, HERSCHEL, PLANCK, GAIA, ALMA, SKA) could be hardly effectively exploited without the VO infrastructure and ideas. All large projects today have to allocate about 30-50% of budget to the processing and data handling software, including the VO interface.

## Basic principles of VO

The VO infrastructure is based on common data format - VOTable.  It is a XML document describing the structure of a data table, the names and physical meaning of data variables (so called meta-data)  using Universal Contents Descriptors (UCD) and data itself in form of ASCII or binary table and/or FITS file [http://vizier.u-strasbg.fr/doc/VOTable] .

The important aspect of VO - its orientation to (locally provided) services requires the dynamically updated list of available services, their descriptions and servers providing these services.  This is provided by the distributed network of VO registries, working like the Domain Name Service servers (DNS) for Internet. Some of the URLs look like ivo://provider/collections.

For seamless exchange of data the VO developers were inspired by commercial business to business (B2B) infrastructure of Web Services (WS) using the protocol SOAP to exchange information in common format similar to HTTP transfered XML files. The service parameters are described by Web Services Description Language (WSDL).  For access to most of the data, several VO protocols have been invented [www.ivoa.net/Documents] :

**ConeSearch** - for searching the vicinity of given position on sky in catalogues. It returns the VOTable with list of objects contained in circle (or square) of given size around particular coordinates.

**Simple Image Access protocol (SIAP)** returning the  images in given format of  given size of objects at given coordinates.

**Simple Spectra Access Protocol (SSAP)** returns spectra in given spectral range of object at given coordinates

**Simple Line Access Protocol (SLAP)** returns list of spectral lines from various line-lists placed around the particular wavelength/energy or in given spectral range.

For messaging of transient events (microlensing event, gamma ray burst, exoplanet transit) the XML message called **VOEvent** is replacing the astronomical telegrams. It has been used mainly for quick follow-up observations of the event by robotic telescopes as it is well-structured and thus machine readable.

For querying of archives the special language **ADQL** (Astronomical Data Query Language) has been developed which allows to limit the search to particular region around given position using word REGION. Most important is the keyword XMATCH for cross matching several catalogues with given level of statistical confidence (in sigmas).

## VO-compatible applications

There is a couple of applications understanding various Virtual Observatory protocols or visualizing VOTables [http://www.ivoa.net/twiki/bin/view/IVOA/IvoaApplications] . Their number has been continuously increasing and more and more astronomical tools have been developed already with VO in mind, or the VO standards have been added to the legacy applications.

Probably the most interesting for amateurs is the **Aladin** applet and Java applications [http://aladin.u-strasbg.fr] recently augmented by **VOSpec** [http://esavo.esa.int/vospec] for preview of spectra. Quite important for analysis of spectra are also **SPLAT-VO** [http://star-www.dur.ac.uk/~pdraper/splat/splat-vo] and the quite complex analysis tool **SpecView** [http://www.stsci.edu/resources/software_hardware/specview] .

The identification of unknown sources in given sky image and estimating their magnitudes can be made easily with **WESIX** (a web based interface to famous SExtractor with cross matching interface). The user can even upload his own image [http://nvogre.phyast.pitt.edu:8080/wesix] for the private analysis.

## New science with VO

The VO infrastructure allows not only work with huge distributed data sets (using the cross matching of large catalogues), but provides the efficient tools for multi-wavelength research. Thus a new yet unknown types of objects may be found (e.g the obscured quasars – Padovani et al. 2004) or very rare objects that escape the attention looking only in one spectral range, e.g. the post-AGB behaviour of stars - Tsalmantza et al.(2006) or brown dwarfs of L and T class (Solano et al. 2006). Another example of VO power is the **VOSED** – the Spectral Energy Distribution Builder that can estimate the bolometric magnitude of the object by direct integration of fluxes obtained by different instruments (from radio to gamma) [http://sdc.laeff.inta.es/vosed/index.jsp] .

## Theoretical Virtual Observatory (TVO)

The current theoretical simulations of galaxy or cluster evolution or magneto-hydrodynamic models of plasma flows around AGN produce terabytes of data comparable by nature to the real observations. So it was quite obvious to use the VO infrastructure for handling this data. There is even the concept of the **Virtual Telescope** which "observes" the results of simulation with given blurring and PSF of a real telescope and then sophisticated methods of VO can be used to search for the similar image in real observation archive of given telescope. (see [http://cds.u-strasbg.fr/twikiDCA/pub/EuroVODCA/KickOff/Trieste_presentation.ppt] or [http://www.laeff.inta.es/svo/Laeff/svo/Cervino_IA.ppt] ).

## Is VO the astronomy without telescopes?

The current astronomy has been gradually changing the concept of the astronomical observation. The most common practice today - writing a proposal, going to the telescope, observing there with a assistant, getting data, bring it home, reduce it, analyze and publish - will probably change soon mostly in the pattern: look at web using VO tools, play with data, analyze it, publish. Of course, the most important is to have the bright idea about the relations of some objects in Universe and ask the interesting questions pushing the knowledge of the mankind further. There is a number of aspects foreseeing the changes in the organization of observations:

9) More and more telescopes today are (at least partially) switching to service observation (executed by the staff on behalf of principal investigator, but without his/her presence)

10) Many telescopes operate in assisted (Keck) or full remote mode (e.g. Palomar P60 [http://www.astro.caltech.edu/~derekfox/P60]), there is a growing number of robotic telescopes (Liverpool 2m at La Palma [http://telescope.livjm.ac.uk] ) even for spectroscopy observation (TSU 2m

[http://schwab.tsuniv.edu/t13.html] , STELLA [http://www.aip.de/stella] ) according to given schedule .

11) Some large telescopes can work only according to fixed schedule due to their construction - e.g. SALT [http://www.salt.ac.za] or HET [http://www.as.utexas.edu/mcdonald/het/het.html] .

12) The quality of observation cannot be estimated immediately without going to some quick-look pipeline (e.g. interferometry, 3D spectroscopy). Even for the final reduction the astronomer has to trust the automatic pipelines, he cannot spend half year learning all the tricky behaviour of given instrument and setting the couple of parameters in various reduction tasks. Moreover, the pipelines provide the homogeneous sets of equally good (or bad) quality without the personal bias.

13) Almost all sources of massive data sets are publicly available after some short proprietary period (1 or 2 years) or immediately. There are projects expecting the data to go to public immediately after reduction (e.g. LSST [http://www.lsst.org] ).

Summarizing all together we can imagine that most observations from future telescopes and satellites will go to the automatic pipelines and the results will be quickly available for anybody in VO. There is, however, still difference in attitude between survey telescopes and pointed observation (where usually the long proprietary period is applied), but this is expected to change as there is a lot of justifications confirming that collaborative effort in science is more rewarding for all participants than personal benefits of private usage of data - sitting on data, sometimes referred to as Data Jealousy .

## VO - the opportunity for everyone

The important aspect of free data availability together with efficient VO infrastructure and sophisticated tools is the opportunity to overcome the phenomenon of Digital Divide [http://en.wikipedia.org/wiki/Digital_divide] . It says roughly that the economically strongest countries preserve their hegemony through the access to the advanced (mostly digital) technologies. Similar problems in astronomy (e.g. access to largest telescopes) may be circumvented by public available VO-compatible archives of these telescopes.

But it also means that an experienced amateur astronomer or student of astronomy with bright ideas may conduct important astronomical research at a professional level using the equal tools and data as the professional astronomer at highly-ranked institution.

## VO and the astronomical community

The VO has already proved itself as a useful tool of current astronomical research but it is considered to be indispensable for the research conducted with future giant telescopes (and satellites). It has been gaining the wide acceptance of major funding agencies (NVO has been funded by NSF, ASTROGRID by PPARC, EURO-VO project by EU FP6) but also in the general astronomical community. At the IAU General Assembly in Prague (2006) there had been several days devoted to the examples of scientific exploitation of current VO and to the questions of astronomical data management in general. Most important issues were articulated in the Astronomer's Data Manifesto [http://arxiv.org/pdf/astro-ph/0701361] .

## Conclusion

Despite many (yet unsolved) issues of the Virtual Observatory (e.g. privacy/openness of astronomical archives, the data quality control, problems of unique semantic descriptions) the VO has been in active development, it is gaining massive support and seem to be the efficient tool for coping with the data avalanche threatening otherwise to bury future astronomical research in the huge data mess. If used properly in open collaborative environment, it can boost our knowledge of universe considerably, especially in fields requiring large-scale statistics, investigation of extremely rare classes of objects, synoptic and multispectral research as well as comparison of huge databases of theoretical models with observations. The VO allows the data mining of distributed databases limited only by the phantasy of the researcher. So it is a excellent opportunity for involvement of amateur astronomers in the professional research as well.

## Acknowledgments

This work was supported by grant GACR 205/06/0584 and EURO-VO DCA WP6.

## References

Padovani, P. et al. 2004, Astron. Astrophys., 424, 545
Richards, A. et al. 2005, Memorie della Societa Astronomica Italiana, 76, 467

Solano, E. et al.  2006, The Virtual Observatory in Action: New Science, New Technology, and Next Generation
       Facilities, in proceedings of 26th meeting of the IAU, Special Session 3,  August 2006 in Prague, Czech
       Republic,  SPS3,  #29

Tsalmantza, P. et al.  2006, Astron. Astrophys., 447, 89